\documentclass[aps,prb,twocolumn,superscriptaddress,longbibliography]{revtex4-2}

\usepackage{graphicx,amsmath,gensymb}
\usepackage[colorlinks=true,allcolors=blue]{hyperref}

\begin{document}

\title{Dielectric Screening in Electromagnetic Dressing of Semiconductors}
\author{Quentin Courtade}
\email{quentin.courtade@u-bordeaux.fr}
\author{Umberto Dellasette}
\email{umberto.dellasette@u-bordeaux.fr}
\author{Sotirios Fragkos}
\author{Stéphane Petit} 
\author{Dominique Descamps}
\author{Yann Mairesse}
\author{Samuel Beaulieu}
\email{samuel.beaulieu@u-bordeaux.fr}

\affiliation{Universit\'e de Bordeaux - CNRS - CEA, CELIA, UMR5107, F33405 Talence, France}

\begin{abstract}
Nonequilibrium manipulation of quantum materials via electromagnetic dressing provides an on-demand route to tailoring electronic band structures through Floquet engineering. Time- and angle-resolved photoemission spectroscopy offers a direct means to probe these light-dressed electronic states. In such photoemission experiments, electromagnetic dressing can also occur for quasi-free electrons outside the material, giving rise to so-called Volkov states. In certain cases, most notably in metallic systems, strong surface screening reduces the penetration of the driving electromagnetic field into the solid, resulting in Volkov contributions that dominate over Floquet ones. This underscores the importance of carefully accounting for the material’s dielectric response when analyzing the origin of light-induced band replicas using time- and angle-resolved photoemission spectroscopy. In this work, we systematically investigate the influence of materials' dielectric properties on Floquet-Volkov dressing of semiconductors, focusing on bulk layered van der Waals materials GeS, SnS, and 2H-WSe$_2$. First, by combining a simple model based on Fresnel equations with an electron-scattering description of Volkov amplitudes, we use polarization-dependent Volkov sideband intensities to extract a lower bound for the real part of the materials' dielectric function. The extracted values typically lie between the reported dielectric constants for monolayer and bulk crystals. Furthermore, we demonstrate that increasing the pump fluence of the infrared pump laser enables the generation of high-order Volkov sidebands, which exhibit clear signatures of nonlinear light–matter interactions in their temporal dynamics, polarization dependence, and angular distributions. Finally, we show that for our experimental geometry, the quasi-transparent nature of semiconductors in the below-band-gap driving regime allows the optical pump to propagate within the sample and undergo multiple total internal reflections. This process produces a series of temporally delayed Volkov replicas in pump–probe measurements via electromagnetic dressing of photoelectrons by the evanescent field associated with total internal reflections. Together, these systematic studies uncover several previously unexplored aspects of Floquet–Volkov dressing in solids, highlighting the central role of dielectric screening of the driving field.
\end{abstract}

\maketitle

\section{Introduction}
Driving materials with electromagnetic fields provides nonequilibrium pathways to access and manipulate quantum states of matter~\cite{Basov17, Torre21, bao_light-induced_2021}. Among these approaches, Floquet engineering has emerged as one of the most promising routes for the optical control of materials out of equilibrium, allowing to tailor their transient properties through hybrid light-matter band structure engineering. This technique relies on the dressing of Bloch states by time-periodic photonic perturbations, giving rise to Floquet-Bloch states whose properties can be tuned by controlling the characteristics of the driving light. Indeed, the remarkable success of Floquet engineering stems from the high degree of control over the periodic driving field, including its intensity, frequency, and polarization. This versatility enables precise manipulation of light-dressed quantum states of matter \cite{Oka09, Wang13, Sie15, mahmood_selective_2016, Oka19, Rudner20, Reutzel20, McIver20, Aeschlimann21, zhou_floquet_2023, zhou_pseudospin-selective_2023, bielinski_floquetbloch_2025, Kobayashi23, liu23, weitz24, Bao2024, Ito23, neufeld_band_2023, fragkos_floquet-bloch_2025}. Time- and angle-resolved photoemission spectroscopy (trARPES) allows direct access to the electronic band structure of solids out of equilibrium~\cite{Wang13, mahmood_selective_2016, Ito23, zhou_floquet_2023, zhou_pseudospin-selective_2023, choi_observation_2025, merboldt_observation_2025, Bao2024, bielinski_floquetbloch_2025, neufeld_band_2023, fragkos_floquet-bloch_2025}. Through its ability to resolve the energy, momentum, and time evolution of photoemitted electrons, trARPES has emerged as a tool of choice for investigating Floquet physics. While Floquet–Bloch states reflect the coherent dressing of electronic bands by the periodic light-field, the detected signal in photoemission experiments also involves quasi-free electrons propagating outside the material, which can themselves be dressed by the driving field (optical pump beam), giving rise to so-called Volkov states (see \textit{e.g.}~\cite{mahmood_selective_2016}). Moreover, since Floquet and Volkov transitions end at the same final energy and in-plane momentum, quantum-path interferences entangle the two contributions in trARPES experiments. 

These competing light-matter dressings (Floquet and Volkov) crucially depend on the driving light polarization state as well as on the dielectric environment at the surface (see \textit{e.g.}~\cite{park_interference_2014, mahmood_selective_2016, choi_observation_2025, merboldt_observation_2025, bao_floquet-volkov_2025,fragkos_floquet-bloch_2025, gadge_comparative_2026}). The dielectric function governs both the penetration depth and the relative strength of the electric field components, thus shaping the relative weight between Floquet and Volkov channels. Hence, understanding how dielectric properties shape the driving field is of primary importance in disentangling the Floquet-Bloch and Volkov amplitudes in trARPES experiments. 

In metallic systems, the formation of Floquet–Bloch states is highly suppressed in favor of Volkov states, since they exhibit strong electronic screening and low penetration depth of the driving field. One pioneering work from Keunecke \textit{et al.} (2020)~\cite{keunecke_electromagnetic_2020} investigated the role of dielectric screening on the formation of light-induced sidebands in Au(111) in trARPES experiments. They proposed a polarization- and momentum-resolved model of the Volkov contribution to the first-order sideband inspired by the work of Park~\cite{park_interference_2014}, for two limiting cases. The first case is when the metal totally screens the electric field, \textit{i.e.}, when the incident electric field is totally reflected by the metallic surface, resulting in a vanishing in-plane electric field on the sample; the second case is when the field is not screened, \textit{i.e.}, when the in-plane electric field on the sample is equal to the in-plane components of the incident electric field. Experimentally, to study the influence of the driving field polarization state on the formation of sidebands, the authors continuously changed the linear polarization axis of the pump laser beam between s- and p-polarized light. They observed that the intensity of the sideband is maximized for a p-polarized pump and minimized for an s-polarized pump. Theoretically, they showed that for the fully screened case, the momentum distribution of the Volkov intensity is maximized at zero momentum (at $\Gamma$) and decreases isotropically and cosinusally as the outgoing electrons' momentum increases. Furthermore, in this fully screened case, the Volkov intensity completely vanishes for an s-polarized pump.
In the unscreened case, however, the momentum distribution of the Volkov intensity becomes more complex. It exhibits some additional (forward-backward and up-down) anisotropies in momentum space. Moreover, the Volkov intensity is more intense than in the fully screened case and is non-zero for an s-polarized pump beam at high momenta. Their experimental data are well reproduced by the model when total screening is assumed. This implies that, in a metal, the strong screening of the driving field at the surface leads to a short penetration depth, implying that the replicas mostly come from the Volkov mechanism. Therefore, in this limiting case of strong screening, the electromagnetic dressing of Bloch electrons to form Floquet-Bloch states is not efficient.

More recently, Wenthaus \textit{et al.}~\cite{wenthaus_insights_2024} reported a strong material dependence in laser-assisted photoemission (Volkov) from time-resolved X-ray core-level spectroscopy. They studied laser-assisted photoemission under identical driving field parameters, in Pt(111) and W(110). They found sideband generation up to the sixth order in W(110) and up to the fourth order in Pt(111). The different response has been attributed to the electronic screening and the resulting modification of the spatial distribution of the field at the surface, leading to a four-fold enhancement of the field in W(110) with respect to Pt(111). These recent findings demonstrate that the Volkov sidebands' intensity is dominated by the electric field experienced by the photoelectrons on the surface, thus highlighting the central role of material-specific dielectric properties in shaping the surface field intensity and distribution. While Ref.~\cite{wenthaus_insights_2024} focused on momentum-integrated core-level photoemission, these findings suggest that even more subtle signatures of dielectric screening can be accessed by analyzing the momentum-resolved distribution of Volkov sideband intensities.

Beyond metals, where strong screening leads to dominant Volkov dressing over Floquet engineering, reduced screening in topological insulators, semimetals, and semiconductors enables the emergence of genuine and efficient light-induced Floquet band dressing. The first experimental evidence of Floquet-Bloch states and associated band renormalization observed using trARPES was reported on Bi$_2$Se$_3$ by Wang \textit{et al.}~\cite{Wang13}. Later, this material class also served as a platform to demonstrate Floquet-Volkov interferences~\cite{mahmood_selective_2016}, Floquet–Bloch manipulation of the Dirac gap in a topological antiferromagnet~\cite{bielinski_floquetbloch_2025}, and the build-up and dephasing of Floquet-Bloch bands on sub-optical-cycle timescales~\cite{Ito23}. 

Quantum path interference between Floquet-Bloch and Volkov states has also recently been used as a hallmark for the first photoemission demonstration of the emergence of Floquet states in graphene~\cite{merboldt_observation_2025,choi_observation_2025}, almost two decades after its theoretical prediction by Oka and Aoki~\cite{Oka09}. In a very recent follow-up paper on this topic, Gadge and coworkers~\cite{gadge_comparative_2026} studied in detail Floquet-Volkov dressing of graphene, showing how photoemission matrix elements, polarization, incidence angle, and near-surface screening shape the momentum-resolved sideband intensity observed in trARPES using advanced theoretical methods, \textit{i.e.}, first-order perturbative Born approximation and time-dependent non-equilibrium Green’s functions.

A significant body of recent literature on Floquet engineering probed by trARPES has focused on the anisotropic semiconducting material black phosphorus (BP). Initial studies demonstrated pseudospin-selective Floquet band engineering~\cite{zhou_pseudospin-selective_2023}, in which pseudospin selectivity emerges when the resonant pump light polarization is aligned with specific high-symmetry directions of BP, thereby enforcing optical selection rules dictated by the material’s pseudospin texture. Following this work, subsequent studies investigated Floquet engineering of BP under below-band-gap pumping~\cite{zhou_floquet_2023}, Floquet-Volkov interference effects~\cite{bao_floquet-volkov_2025}, and Floquet-induced glide-mirror symmetry breaking~\cite{Bao2024_glide}. Moreover, two recent studies reported symmetry-guided selection rules in Floquet engineering of BP, demonstrating that specific light polarization aligned along a specific crystal direction can be used to manipulate the symmetry (parity) of Floquet-state wavefunctions~\cite{Bao2024, fan_floquet_2025}.

Floquet-Bloch states were also recently observed in semiconducting transition metal dichalcogenides (TMDCs)~\cite{Aeschlimann21, fragkos_floquet-bloch_2025}. In these systems, the broken inversion symmetry gives rise to non-vanishing and opposite Berry curvatures in the K and K' regions of the Brillouin zone, which is well-known to lead to chiroptical selection rules for resonant interband transitions (see \textit{e.g.}~\cite{Mak12, Zeng12, Cao12, beaulieu_berry_2024}). These selection rules are at the heart of valleytronics. Extending these ideas to Floquet engineering, Fragkos \textit{et al.} demonstrated the formation of valley-polarized Floquet-Bloch states in 2H-WSe$_2$ upon below-band-gap driving with circularly polarized light pulses~\cite{fragkos_floquet-bloch_2025}. Their results also reveal the signature of quantum-path interference between Floquet-Bloch and Volkov states in TMDCs. Finally, they conclude from extreme ultraviolet photoemission circular dichroism that Floquet engineering allows for controlling the orbital character of hybrid light-matter states~\cite{fragkos_floquet-bloch_2025}. 

These results in semiconducting 2H-WSe$_2$ and black phosphorus under below-band-gap pumping highlight the fundamental interest of this regime for Floquet engineering. Under these conditions of light-matter interaction, the incident field is partially screened and reshaped at the surface by the dielectric environment, the material is quasi-transparent, and the optical penetration depth becomes large. Our work aims to clarify the role of the material’s dielectric response in shaping the observed photoemission features associated with electromagnetic dressing of semiconductors. We show that, in this regime, the polarization dependence of Volkov sideband intensity depends on the materials' dielectric properties, enabling the extraction of a lower bound for the real part of the dielectric function. We further show that the strong driving fluence typically required for efficient Floquet engineering leads to the emergence of high-order Volkov sidebands, whose temporal profile, angular distributions, and driving polarization state dependencies are strongly reshaped by the nonlinear nature of the light-matter interaction. Finally, we demonstrate that the combination of below-band-gap excitation and large incidence angle gives rise to multiple temporally delayed Volkov sidebands originating from total internal reflections of the pump field inside the crystal. This phenomenon is unique to semiconductors driven in their transparency windows and provides additional insight into the dielectric screening in electromagnetic dressing of semiconductors.

\section{Results and discussion}

\subsection{Experimental setup}

\begin{figure}
    \centering
    \includegraphics[width=\linewidth]{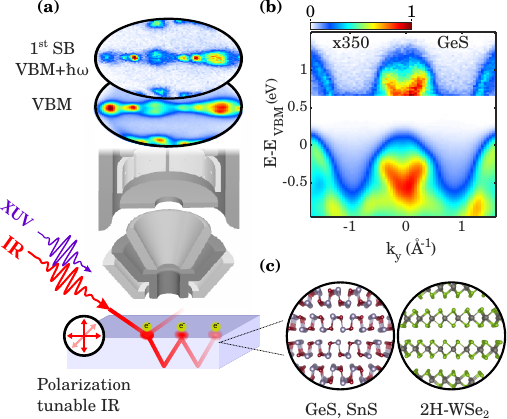}
\caption{\textbf{Schematic of the experimental setup.} \textbf{(a)} A polarization-tunable infrared (IR) pump (1.2~eV, 135~fs) and a p-polarized extreme ultraviolet (XUV, 21.6~eV) probe pulses are focused on the sample in the interaction chamber of a time-of-flight momentum microscope at an incidence angle $\theta=65^\circ$. We show constant energy contours measured for the semiconductor GeS, for the incidence plane aligned along the mirror plane, integrated over all pump polarizations, at pump-probe temporal overlap, at the energies of the valence band maximum (VBM) and of the first-order sideband ($\mathrm{VBM}+\hbar\omega$). \textbf{(b)} Experimentally measured photoemission intensity from GeS along $k_y$, with the light scattering plane aligned along the armchair direction, summed over all pump polarizations, at pump-probe temporal overlap. \textbf{(c)} Schematic of the crystal structures of GeS, SnS, and 2H-WSe$_2$.}
    \label{fig:figure1}
\end{figure}

We performed time- and angle-resolved photoemission spectroscopy (trARPES) on GeS, SnS, and 2H-WSe$_2$ (see the experimental scheme in Fig.~\ref{fig:figure1}(a)). The bulk samples are commercial samples bought from \textsc{HQ Graphene}. Their lateral size is 2-3~mm~$\times$~2-3~mm. They are mounted on a flag-style sample holders using a UHV-compatible conductive epoxy glue and are cleaved \textit{in-situ} under ultra-high vacuum conditions by mechanically striking a cylindrical ceramic post glued onto the sample. The freshly cleaved samples are then introduced into a motorized hexapod for sample alignment inside the main chamber. All experiments were conducted at room temperature. More details about the experimental preparation can be found in~\cite{Fragkos2025,tkach_multimode_2026}.
Our beamline is pumped by an amplified Yb fiber laser delivering 50~W average power at a repetition rate of 166~kHz, with a central photon energy of 1.2~eV (1030~nm) and a sub-150~fs pulse duration. In the probe arm, we first frequency-double the fundamental beam in an $\alpha$-BBO crystal (515~nm). To eliminate the 515~nm driven laser beam from the high-order harmonic extreme ultraviolet (XUV) beam, the green beam is spatially shaped into an annular beam before being focused into a dense argon gas jet. The 515~nm driver is spatially filtered after the gas jet using a pinhole. After XUV generation through high-order harmonics generation (HHG), the 9$^\mathrm{th}$ harmonic of the frequency-doubled fundamental (21.6~eV) is spectrally selected using a combination of multilayer mirrors and an Sn filter. The pump (IR, 1.2~eV) and probe (XUV, 21.6~eV) pulses are recombined and focused onto the sample in the interaction chamber of a time-of-flight momentum microscope, at an incidence angle $\theta = 65^\circ$~\cite{Fragkos2025}. The linear polarization axis of the XUV probe pulses can be tuned by rotating a half-wave plate placed in the 515~nm beam immediately before the HHG chamber. Similarly, the linear polarization axis of the infrared (IR) pump pulses can be continuously adjusted by rotating a half-wave plate located just upstream of the recombination chamber. The energy- and momentum-resolved photoemission intensities, measured as a function of pump-probe delay or pump polarization state, are acquired using a custom time-of-flight momentum microscope~\cite{TKACH2025114167}. Our setup has a temporal resolution of 144~fs~\cite{Fragkos2025}. More details about our apparatus can be found elsewhere~\cite{Comby22, Fragkos2025}. The light-scattering plane is aligned along selected high-symmetry directions of the samples (armchair or zigzag for GeS and SnS, $\Gamma$-M for 2H-WSe$_2$, see Fig.~\ref{fig:figure1}(c)).
As a representative result, Fig.~\ref{fig:figure1}(b) shows the experimental band structure of GeS along the $k_y$ axis, at pump-probe temporal overlap, integrated over all IR pump polarizations (continuous rotation of a half-wave plate in the pump beam during the measurement) with a pump IR fluence of $\sim$1.6~mJ/cm$^2$. A replica of the valence band (VB) appears at an energy upshifted by one pump photon energy, corresponding to the light-field-dressed VB (see the associated constant-energy contours in Fig.~\ref{fig:figure1}(a)).

\subsection{Polarization-resolved momentum distribution of Volkov sideband intensities}
\label{sec:alpha}
In this section, we introduce the revised Fresnel-Volkov model to quantitatively express the Volkov intensity as a function of outgoing photoelectron momentum, light polarization state, and the material's dielectric constant. To do so, we start by assuming an incident linearly polarized driving IR field $\textbf{E}(t)=\textbf{E}_i\cos(\omega t)$ whose polarization state is described by the polarization angle $\phi$, at an angle of incidence $\theta$ with the sample. As shown by Keunecke and coworkers~\cite{keunecke_electromagnetic_2020} and Gadge and coworkers~\cite{gadge_comparative_2026}, the amplitude $\textbf{E}_i$ of the incident driving field on the sample surface reads:
\begin{equation}
    \textbf{E}_i=E_0\begin{pmatrix}
        \cos\theta\cos\phi \\ \sin\phi \\ \sin\theta\cos\phi \\
    \end{pmatrix}.
    \label{eq:Ei}
\end{equation}
Inspired by~\cite{park_interference_2014, gadge_comparative_2026} and~\cite{fragkos_floquet-bloch_2025}, we use Fresnel equations to describe the total electric field in the vicinity of the surface of the sample $\textbf{E}^\mathrm{IR}$, as a function of the dielectric constant of the material $\epsilon$. Assuming that the material is not magnetic and does not absorb upon below-band-gap pumping, the dielectric constant is linked to the refractive index $n$ by $n=\sqrt{\epsilon}$ and takes real values.
This results in an effective field $\textbf{E}^\mathrm{IR}$ which has contributions from both the incident and the reflected fields. It reads:

\begin{equation}
    \textbf{E}^\mathrm{IR}=\begin{pmatrix}
        E_x^i-E_x^r \\ E_y^i+E_y^r \\ E_z^i+E_z^r \\
    \end{pmatrix} \underset{\mathrm{Eq}.\ref{eq:Ei}}{=} E_0 \begin{pmatrix}
        \cos\theta\cos\phi(1-r_p) \\ 
        \sin\phi(1+r_s) \\
        \sin\theta\cos\phi(1+r_p) \\
    \end{pmatrix},
    \label{eq:Etot}
\end{equation}
with $r_s$ and $r_p$ the Fresnel coefficients for the s- and p-polarized components of the driving field, which can be expressed as functions of $\epsilon$:
\begin{align}
    r_s&=\frac{\cos\theta-\sqrt{\epsilon-\sin^2\theta}}{\cos\theta+\sqrt{\epsilon-\sin^2\theta}},\label{eq:fresnel_coeffs_rs} \\
    \ r_p&=\frac{\epsilon\cos\theta-\sqrt{\epsilon-\sin^2\theta}}{\epsilon\cos\theta+\sqrt{\epsilon-\sin^2\theta}}.
    \label{eq:fresnel_coeffs_rp}
\end{align}

Then, to describe the momentum- and polarization-resolved Volkov sideband intensity, we use the model proposed by Keunecke \textit{et al.}~\cite{keunecke_electromagnetic_2020} and inspired by Park's model~\cite{park_interference_2014}. The intensity of the first-order sideband $I_1(\textbf{k},\phi)$ is given by~\cite{keunecke_electromagnetic_2020}:

\begin{equation}
    I_1(\textbf{k},\phi)\sim I_0(\textbf{k},\phi)\times|a_1(\textbf{k},\phi)|^2,
    \label{eq:I1}
\end{equation}
where $I_0(\textbf{k},\phi)$ is the valence band intensity and $|a_1|^2$ is the sideband amplitude, which has contributions from both Volkov ($\alpha$) and Floquet ($\beta$) transitions~\cite{keunecke_electromagnetic_2020}:

\begin{equation}
    |a_1|^2\sim\frac{1}{4}(\beta-\alpha)^2.
    \label{eq:a1}
\end{equation}
At a given angle of incidence, the Volkov amplitude $\alpha$ then depends on the photoelectron momentum \textbf{k} and on the polarization angle of the driving beam $\phi$. It reads~\cite{keunecke_electromagnetic_2020}:

\begin{equation}
    \alpha=\frac{e}{m_e\omega^2}\textbf{E}^\mathrm{IR}\cdot\textbf{k},
    \label{eq:alpha}
\end{equation}
where $m_e$ is the mass of the electron, $e$ the electric charge and $\omega/2\pi$ is the driving frequency. This allows us to rewrite $\alpha$ as an explicit function of the dielectric constant, in a similar form as in~\cite{fragkos_floquet-bloch_2025, gadge_comparative_2026}:

\begin{equation}
    \alpha(\textbf{k},\theta,\phi,\epsilon)=\frac{eE_0}{m_e\omega^2}\hat{\epsilon}(\theta,\phi,\epsilon)\cdot\textbf{k},
    \label{eq:alpha_revised}
\end{equation}
with:
\begin{equation}
    \hat{\epsilon}(\theta,\phi,\epsilon)=\begin{pmatrix}
        \cos\theta\cos\phi(1-r_p) \\
        \sin\phi(1+r_s) \\
        \sin\theta\cos\phi(1+r_p)
    \end{pmatrix}.
    \label{eq:epsilon_hat}
\end{equation}
This is what we call the revised Fresnel-Volkov model, which now explicitly expresses the distribution of the Volkov intensity as a function of electron momentum, light polarization state, and materials' dielectric constant.

As discussed in Refs.~\cite{keunecke_electromagnetic_2020, wenthaus_insights_2024,gadge_comparative_2026}, the distribution of the electric field at the surface plays a central role in Volkov and Floquet dressings. Material-specific dielectric properties reshape the near-surface field distribution, thus modulating both Floquet and Volkov intensities.
To further highlight the difference between highly screening metals and partially screening semiconductors, we calculate the electric field components from light-driven Au(111) following Ref.~\cite{keunecke_electromagnetic_2020} and of the semiconductors investigated in this study, GeS and 2H-WSe$_2$, using Eq.~\ref{eq:Etot}. 

\begin{figure}[t!]
    \centering
    \includegraphics[width=\linewidth]{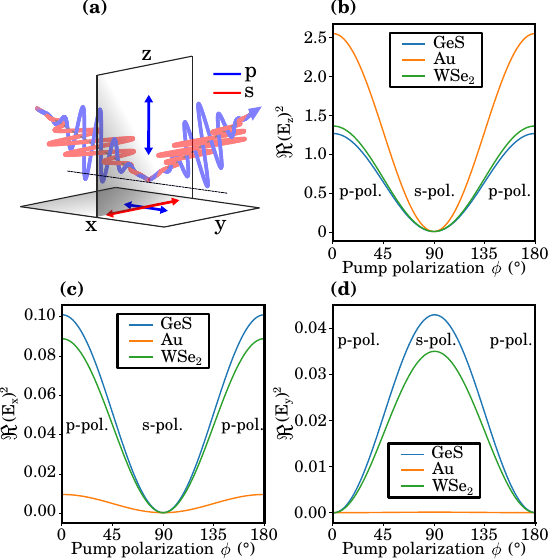}
    \caption{\textbf{Simulated polarization-dependent field components for Au(111), GeS and 2H-WSe$_2$ for an incidence angle of 65$^\circ$.} \textbf{(a)} Schematic representation of the experimental geometry and of the partially reflected electric field with reference to our axis convention. The two in-plane components are $x$ and $y$. When the pump polarization is completely in-plane (s-polarization), the electric field is along $y$. \textbf{(b-d)} Square of the real part of the electric field ($\Re{(\mathbf{E})}^2$) of Au(111) (orange), GeS (blue), and 2H-WSe$_2$ (green) projected along $z$ (b), $x$ (c), and $y$ (d), calculated through Eq.~\ref{eq:Etot} as a function of the pump linear polarization angle ($\phi$).}
    \label{fig:field_component}
\end{figure}

A schematic representation of the experimental geometry is shown in Fig.~\ref{fig:field_component}(a). The laser impinges on the sample at an incidence angle $\theta = 65^\circ$ and is partially reflected at the surface. In our convention, the $y$-axis denotes the in-plane direction of the s-polarized pump beam (while a p-polarized pump lies within the $zx$-plane). For different materials, Figs.~\ref{fig:field_component}(b-d) show the squared value of the real part of each field component for each linear pump polarization angle ($\phi$). 
If the material is absorbent, then the refractive index $\tilde{n}$ is complex-valued and reads: $\tilde{n}=n+ik$. From now on, quantities denoted with a tilde are complex-valued.
The values of $n$ and $k$ used for Au(111) are $n=0.189$ and $k=6.87$~\cite{C8NR09038F}, for GeS are $n=3.8$ and $k=0$ \cite{al-basheer_determining_2024} and for 2H-WSe$_2$ are $n=4.23$ and $k=1.7\times10^{-5}$ \cite{munkhbat_optical_2022}.
Consequently, the Volkov mechanism will dominate the emergence of sidebands from metals, as the enhanced out-of-plane $E_z$ component near the surface (as shown in Fig.~\ref{fig:field_component}(b)) strongly couples to quasi-free outgoing electrons. For Au(111), one can also see the existence of very low in-plane components $E_x$ and $E_y$. Indeed, the high values of out-of-plane components are related to intense Volkov states, while the almost vanishing in-plane components explain why Floquet dressing is not an efficient mechanism in metals.
In contrast, GeS and 2H-WSe$_2$ exhibit both significant in-plane and out-of-plane field components, enabling the coexistence of Floquet-Bloch and Volkov states. These field distributions thus provide a microscopic explanation for the material-dependent intensity ratio between the two photoemission channels observed across different materials. 

After having established from a theoretical point of view how materials' dielectric properties reshape the field at their surface, we now experimentally study the dependence of the momentum-integrated Volkov sideband intensity ($\alpha$ in Eq.~\ref{eq:a1}) on the pump polarization. Then, by fitting the experimental data using the Fresnel-Volkov model, we will try to extract information about the dielectric properties of the sample under investigation. 

\begin{figure}[t!]
    \centering
    \includegraphics[width=\linewidth]{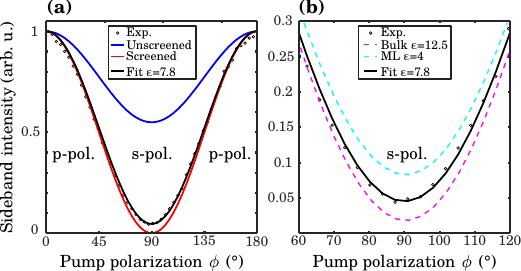}
    \caption{\textbf{Momentum-integrated polarization-resolved sideband intensity in GeS.} \textbf{(a)} Evolution of the momentum-integrated Volkov intensity with the pump polarization axis angle at a pump fluence of $\sim$1.6~mJ/cm$^2$. The black dots are the experimental data, the blue line is calculated in the fully unscreened case, the red curve is calculated in the fully screened case, and the black line is the result from the fit using Eq.~\ref{eq:alpha_revised} ($\epsilon=7.8$). \textbf{(b)} Closer view of (a) around the s-polarized pump ($\phi=90^\circ$). The pink dashed line is the polarization-dependent Volkov intensity calculated using the reported value of the dielectric constant for the bulk material ($11<\epsilon_\mathrm{bulk}<15$~\cite{el-bakkali_layers_2021,koc_mechanical_2015,arfaoui_optical_2023,venghaus_dielectric_1975}; here, the curve is obtained with $\epsilon=12.5$). Similarly, the cyan dashed line is calculated for the monolayer (ML) value ($\epsilon_\mathrm{ML}\sim4$~\cite{el-bakkali_layers_2021}), and the black curve is the result from the fit ($\epsilon=7.8$).}
    \label{fig:eps_fit}
\end{figure}

The results obtained on the semiconductor GeS at a pump fluence of $\sim$1.6~mJ/cm$^2$ are presented in Fig.~\ref{fig:eps_fit}. In Fig.~\ref{fig:eps_fit}(a), the black dots represent the experimental momentum-integrated Volkov sideband intensity as a function of the pump polarization angle $\phi$. This signal has been obtained by taking the normalized ratio between the photoemission intensity from the first-order sideband and the valence band.
To compare the experimental data with the limiting cases presented in~\cite{keunecke_electromagnetic_2020}, we computed in Fig.~\ref{fig:eps_fit}(a) the momentum-integrated Volkov sideband intensity in the fully screened ($E_\mathrm{xy}=0$, in red) and unscreened ($E_\mathrm{xy}\neq0$, in blue) electric field cases. These two extreme cases are computed using the model given in~\cite{keunecke_electromagnetic_2020}, which neglects the reflected field and only considers the incident electric field given in Eq.~\ref{eq:Ei}.
The three curves follow a $\cos^2\phi$ dependence, as predicted in Eq.~\ref{eq:alpha_revised}. As expected, they are maximized for p-polarized pump and minimized for s-polarized pump. However, for the s-polarized pump, the Volkov intensity estimated for the unscreened (screened) case is much higher (lower) than the experimentally measured signal.

In Au(111), the fully screened theoretical model reproduced very well the experimentally measured polarization-resolved Volkov intensity~\cite{keunecke_electromagnetic_2020}. In contrast, for our case in GeS, \textit{i.e.}, a semiconductor under below-band-gap excitation, the experimental polarization-resolved Volkov intensity curve lies between these two limiting cases. Our experimental data can be very well fitted by using the revised Fresnel-Volkov model (black solid line), which incorporates the role of partial screening by the dielectric environment. The fitting of the experimental data is done with a least-squares minimization of the momentum-integrated sideband intensity $|a_1|^2$ from Eqs.~\ref{eq:I1} and~\ref{eq:a1}, where $\alpha$ is calculated using Eq.~\ref{eq:alpha_revised}. The dielectric constant $\epsilon$ is the only free parameter in Eq.~\ref{eq:alpha_revised}, through the Fresnel coefficients (see Eqs.~\ref{eq:fresnel_coeffs_rs} and~\ref{eq:fresnel_coeffs_rp}); $\epsilon$ is then globally estimated.

In Fig.~\ref{fig:eps_fit}(b), we compare the result of our fit with values previously reported in the literature for bulk and monolayer GeS. Our fitted value lies between them: $\epsilon_\mathrm{fit}^\mathrm{GeS}=7.8\pm0.7$, $11<\epsilon_\mathrm{bulk}^\mathrm{GeS}<15$~\cite{el-bakkali_layers_2021,koc_mechanical_2015,arfaoui_optical_2023,venghaus_dielectric_1975} and $\epsilon_\mathrm{ML}^\mathrm{GeS}~\sim~4$~\cite{el-bakkali_layers_2021}. 
We interpret this result by noting that, in an XUV (tr)ARPES experiment, the very short inelastic mean free path of the outgoing photoelectrons implies that the measurement predominantly probes the topmost layer of the sample. As a consequence, the probed material effectively behaves as a monolayer that is bounded by vacuum on one side and by deeper layers of the material on the other, resulting in an intermediate response between the monolayer and bulk limits.
A limitation of our model is the potential presence of Floquet states, which are neglected in our revised description of $\alpha$ and which may influence the polarization-resolved, momentum-integrated sideband intensity. If the contribution of Floquet states were significant in addition to the dominant Volkov states, the momentum-integrated sideband intensity would be expected to increase, in particular for s-polarized pumping, where the Volkov contribution is minimized. In such a scenario, the experimental curve would shift closer to the unscreened limit (see Fig.~\ref{fig:eps_fit}(a)), resulting in a lower fitted value of $\epsilon$. We therefore conclude that the dielectric constants extracted from our analysis represent lower bound estimates for the studied materials.
When artificially adding some enhanced but realistic contributions from competing phenomena such as Floquet states, the fitted dielectric constant remains in the confidence interval of the fitting.
Uncertainties were obtained using standard deviations calculated with a Newey-West estimator.

We now extend the analysis presented above to another crystal from the same material class, namely SnS. By applying the same framework and experimental considerations to SnS, we extract its dielectric constant and benchmark it against known bulk and monolayer values. The fitted $\epsilon$ is $\epsilon_\mathrm{fit}^\mathrm{SnS}=12.2\pm1.6$, which lies between the bulk ($15<\epsilon_\mathrm{bulk}^\mathrm{SnS}<20$~\cite{banai_ellipsometric_2014,nguyen_temperature_2020}) and the monolayer ($\epsilon_\mathrm{ML}^\mathrm{SnS}\sim5$~\cite{batool_dft_2022}) values.
The results of our fittings are summarized in Table~\ref{tab:fittings}.
\begin{table}[b!]
    \centering
    \begin{tabular}{c|c|c} 
            & GeS & SnS \\ \hline
        Bulk & $11<\epsilon<15$~\cite{el-bakkali_layers_2021,koc_mechanical_2015,arfaoui_optical_2023,venghaus_dielectric_1975} & $15<\epsilon<20$~\cite{banai_ellipsometric_2014,nguyen_temperature_2020} \\ 
        Monolayer & $\epsilon\sim4$~\cite{el-bakkali_layers_2021} & $\epsilon\sim5$~\cite{batool_dft_2022} \\ \hline
        Exp. lower bounds & $\epsilon_\mathrm{fit}^\mathrm{GeS}=7.8\pm0.7$ & $\epsilon_\mathrm{fit}^\mathrm{SnS}=12.2\pm1.6$ \\
    \end{tabular}
    \caption{\textbf{Reported and fitted values of the dielectric constants of GeS and SnS.} The reported values are either calculated using DFT and first-principles calculations~\cite{el-bakkali_layers_2021,koc_mechanical_2015,arfaoui_optical_2023,banai_ellipsometric_2014,batool_dft_2022}, or experimental results obtained using electron energy loss spectroscopy~\cite{venghaus_dielectric_1975} or spectroscopic ellipsometry~\cite{banai_ellipsometric_2014,nguyen_temperature_2020}. The values we report (\textit{Exp. lower bounds}) are lower bound estimates that lie between the bulk and monolayer limits, and were obtained using the fitting procedure described in the main text.}
    \label{tab:fittings}
\end{table}

\begin{figure}[t!]
    \centering
    \includegraphics[width=\linewidth]{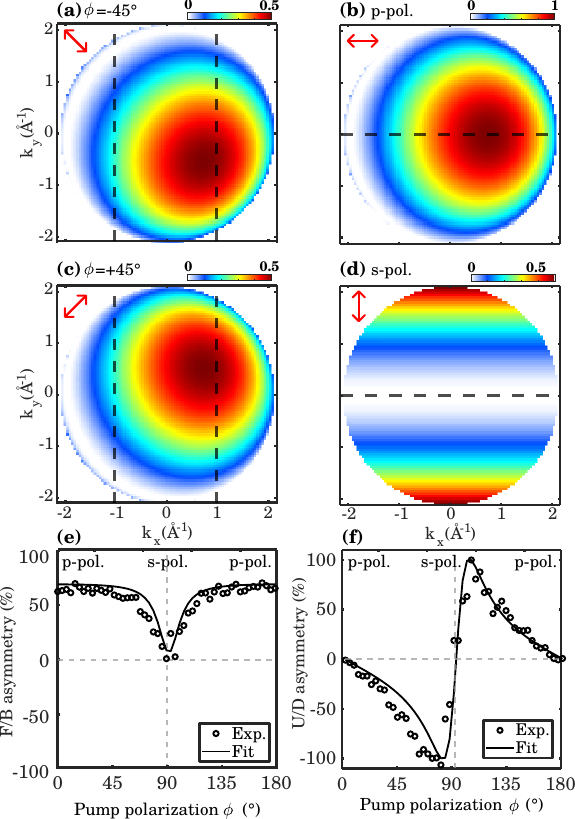}
    \caption{\textbf{Polarization-dependent momentum distribution of the Volkov sideband amplitude in GeS.} \textbf{(a-d)} Momentum-resolved maps of $|\alpha|^2$ calculated using Eq.~\ref{eq:alpha_revised} and the fitted value of $\epsilon=7.8$ (experimental value of GeS) for (a) $\phi=-45^\circ$, (b) $\phi=0^\circ$ (p-polarized pump), (c) $\phi=+45^\circ$, and (d) $\phi=90^\circ$ (s-polarized pump). The red arrows represent the in-plane direction of polarization of the pump. \textbf{(e)} Polarization-resolved forward/backward asymmetry (\textit{i.e.}, along $k_x$) in the momentum-distribution of the first sideband (black dots are experimental results and the black line is the theoretical result using the fitted value of $\epsilon=7.8$). The asymmetry is defined as the normalized difference between the intensity within the forward and backward regions delimited by the dashed black lines in (a) and (c). \textbf{(f)} Polarization-resolved up/down asymmetry (\textit{i.e.}, along $k_y$) of the momentum distribution of the first sideband (black dots are experimental results and the black line is the theoretical result using the fitted value of $\epsilon=7.8$). The asymmetry is defined as the normalized difference between the top and bottom regions delimited by the dashed black lines in (b) and (d).}
    \label{fig:asymmetries}
\end{figure}

Having established the role of dielectric properties in the momentum-integrated Volkov sideband intensity, we go one step further and investigate the polarization-dependent, momentum-resolved distribution of the Volkov sideband intensity. Figs.~\ref{fig:asymmetries}(a-d) show momentum-resolved distributions of $|\alpha|^2$ for four different pump polarizations, calculated using the previously described Fresnel-Volkov model (Eqs.~\ref{eq:a1} and~\ref{eq:alpha_revised}). For $\phi=\pm45^\circ$, the intensity is highly anisotropic, exhibiting pronounced up/down (U/D) and forward/backward (F/B) asymmetries. The direction of the maximum sideband yield follows the direction of the pump polarization. For p-polarized pumping ($\phi=0^\circ$), the intensity is up/down symmetric but forward/backward asymmetric, with stronger intensity in the $+k_x$ hemisphere. For s-polarized pumping ($\phi=90^\circ$), as previously reported \textit{e.g.} by Keunecke and coworkers~\cite{keunecke_electromagnetic_2020}, the Volkov intensity vanishes in the scattering plane (orthogonal to the polarization axis), while remaining finite at large momenta along the pump polarization direction. Using s-polarized driving pulses and detecting photoelectrons in the plane perpendicular to the scattering plane, therefore, constitutes a very efficient experimental strategy to suppress spurious Volkov contributions when searching for pure Floquet physics in trARPES experiments.

To quantitatively analyze polarization-resolved momentum-space redistributions of the Volkov sideband intensity, we extract the forward/backward (Fig.~\ref{fig:asymmetries}(e)) and up/down (Fig.~\ref{fig:asymmetries}(f)) asymmetries for both the experimental data (for GeS) and the corresponding Fresnel-Volkov simulations. These asymmetries are defined as normalized differences between the intensities integrated over the indicated regions of interest in Figs.~\ref{fig:asymmetries}(a-d). As expected, the F/B asymmetry is maximized for p-polarized pumping and minimized for s-polarized pumping, and its value is always positive, as the intensity maximum remains in the same momentum hemisphere for all pump polarizations. As shown in Fig.~\ref{fig:asymmetries}(e), the F/B asymmetry is accurately reproduced by the Fresnel-Volkov simulations using the experimentally extracted dielectric constant. The U/D asymmetry exhibits the same characteristic dependence on the pump polarization as previously reported in~\cite{fragkos_floquet-bloch_2025,choi_observation_2025,merboldt_observation_2025}. It vanishes for both s- and p-polarized pumpings and exhibits a sharp sign flip when going across s-polarization. The experimental data are also qualitatively well reproduced by the simulation employing the fitted dielectric constant. The pump polarization dependence of the F/B and U/D asymmetries in the momentum distributions of Volkov sidebands is expected to be largely universal across investigated material classes. While the dielectric environment can induce changes in the amplitude of the asymmetries and in the slopes characterizing their polarization dependence, the overall trends should remain universal. As shown by Fragkos and coworkers~\cite{fragkos_floquet-bloch_2025}, searching for pronounced deviations from these universal behaviors can serve as an efficient indicator of the presence of Floquet states. Indeed, using circularly polarized driving pulses in 2H-WSe$_2$, they demonstrated that the U/D asymmetry is nonvanishing and reverses sign with the light helicity, a behavior traced back to the emergence of valley-polarized Floquet–Bloch states. This interpretation was further corroborated by time-dependent nonequilibrium Green’s function calculations~\cite{fragkos_floquet-bloch_2025}. This trARPES hallmark of the presence of Floquet states (sitting on a Volkov background) is therefore anticipated to be extendable to other material systems where polarization-controlled Floquet states are expected to emerge.

\begin{figure}[t!]
    \centering
    \includegraphics[width=\linewidth]{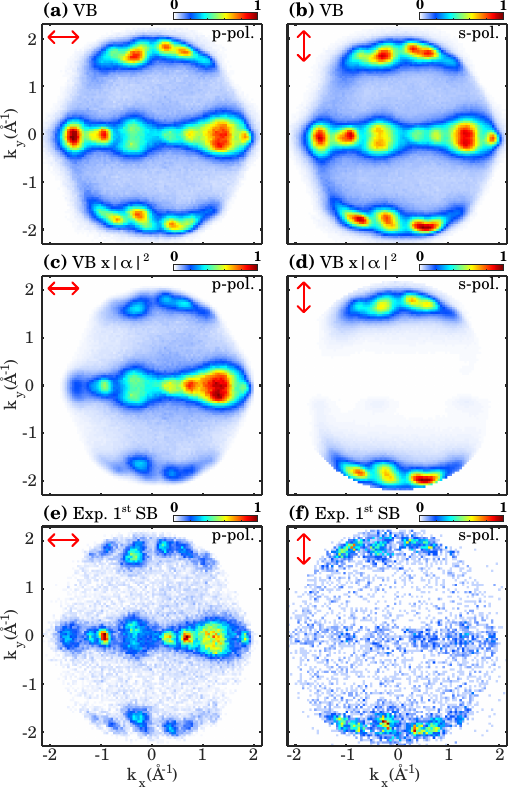}
    \caption{\textbf{Polarization-dependent momentum distribution of photoemission intensity from the valence band (VB) and first-order sideband of GeS, upon below-band-gap pumping at a pump fluence of $\sim$1.6~mJ/cm$^2$.} The red arrows represent the in-plane direction of the pump polarization. \textbf{(a-b)} Experimentally measured momentum distributions of photoemission intensities from valence band ($E=E_\mathrm{VBM}\sim0$~eV) for p- and s-polarized pump, respectively. \textbf{(c-d)} Calculated momentum distribution of photoemission intensities of first-order sidebands obtained from experimental momentum distribution (a) and (b) multiplied by the simulated Volkov sideband intensity using Eq.~\ref{eq:alpha_revised} (with the fitted value of $\epsilon=7.8$). \textbf{(e-f)} Experimentally measured momentum distributions of photoemission intensities from first-order sidebands (integrated over $1.1~\mathrm{eV}<E<1.3~\mathrm{eV}$) for p- and s-polarized pump, respectively.}
    \label{fig:GeS_BM}
\end{figure}

In Fig.~\ref{fig:GeS_BM}, we show the momentum distribution of photoemission intensity from the VB and first-order sideband of GeS, upon below-band-gap pumping, for p- and s-polarized driving pulses. As expected, VB momentum maps are nearly identical for both pump polarizations (p-polarized probe), see Figs.~\ref{fig:GeS_BM}(a) and~(b). The theoretically predicted Volkov sidebands' momentum distributions (see Figs.~\ref{fig:GeS_BM}(c) and~(d)) were obtained by multiplying the experimentally measured VB momentum distributions (Figs.~\ref{fig:GeS_BM}(a) and~(b)) by the simulated $|\alpha|^2$ for p- and s-polarized pumpings. These polarization- and momentum-resolved $|\alpha|^2$ are calculated with the Fresnel-Volkov model using the fitted dielectric constant of GeS, see Eqs.~\ref{eq:I1} and~\ref{eq:a1}. As discussed in the context of Fig.~\ref{fig:asymmetries}(b), for a p-polarized pump, $|\alpha|^2$ is maximized at low outgoing photoelectron momenta (near $\Gamma$), with an enhanced signal in the positive $k_x$ hemisphere. This behavior is evident in Fig.~\ref{fig:GeS_BM}(c), where the theoretically predicted momentum distribution of the Volkov sideband closely resembles the one of the VB, but with an enhanced signal along $k_y=0$ compared to larger $|k_y|$ values. In contrast, for an s-polarized pump, the theoretically predicted momentum distribution of the Volkov sideband exhibits a markedly different behavior: it differs strongly from the VB momentum distribution, showing a vanishing photoemission intensity around $k_y=0$ and a nonzero signal at large $|k_y|$ (see Fig.~\ref{fig:GeS_BM}(d)). This prediction is consistent with the results discussed in the context of Fig.~\ref{fig:asymmetries}(d) and with the conclusions of Keunecke and coworkers~\cite{keunecke_electromagnetic_2020}.

In Figs.~\ref{fig:GeS_BM}(e) and~(f), we present the experimentally measured momentum distributions of the sideband intensities for p- and s-polarized driving pulses, respectively, at a pump fluence of $\sim$1.6~mJ/cm$^2$. Most features observed in these experimental momentum maps are in qualitative agreement with the theoretically predicted momentum distributions of the Volkov sideband shown in Figs.~\ref{fig:GeS_BM}(c) and~(d). Specifically, for a p-polarized pump, the sideband intensity is up/down symmetric and enhanced for positive $k_x$. For an s-polarized pump, the sideband intensity exhibits the expected relative enhancement at large $|k_y|$, accompanied by a strong reduction of the intensity around $k_y=0$. It should be noted, however, that unlike the theoretical expectation (see Fig.~\ref{fig:GeS_BM}(d)), the experimentally measured sideband intensity for an s-polarized pump does not completely vanish around $k_y=0$. The origin of this nonvanishing experimental signal may be twofold. First, it could arise from the presence of Floquet states, which are not included in our Fresnel–Volkov model. Second, it may result from a slight misalignment of the sample, leading to a small out-of-plane component of the electric field at the surface, as recently discussed by Bao and coworkers~\cite{bao_floquet-volkov_2025}. These results further highlight the importance of properly accounting for dielectric properties and polarization state effects when interpreting the momentum-dependent intensity fingerprints of Volkov sidebands, in the spirit of disentangling their signatures from those of Floquet sidebands in trARPES experiments.

\subsection{Higher-order Volkov sidebands}

The intensity of Floquet states and associated band renormalization increase with the driving laser intensity. In trARPES experiments, one therefore typically seeks to increase the pump fluence up to the limit set by pump-induced space-charge effects, in order to enhance the spectral signatures of light-induced Floquet states. Increasing the pump fluence, however, inherently leads to the emergence of nonlinear multiphoton processes, such as unwanted multiphoton photoemission from the pump, but also, more interestingly, the formation of high-order Floquet–Volkov replicas. In this section, we investigate the temporal behavior, polarization dependence, and momentum-space redistribution of photoemission intensity from high-order sidebands, and show that they all carry clear signatures of the nonlinear nature of the light–matter interaction when compared to their first-order counterparts.

\begin{figure}
    \centering
    \includegraphics[width=\linewidth]{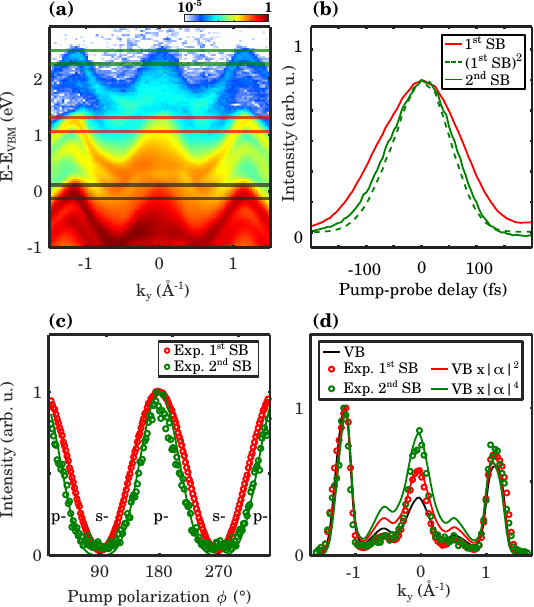}
    \caption{\textbf{Temporal profile, polarization dependence, and angular distribution of first- and second-order Volkov sidebands.} \textbf{(a)} Energy–momentum distribution of photoemission intensity from 2H-WSe$_2$ along K–$\Gamma$–K$^{\prime}$, for a p-polarized XUV probe, at pump–probe temporal overlap, integrated over all IR pump polarizations ($\sim 5.7$~mJ/cm$^2$). The black lines indicate the region of interest for the VB, the red lines for the first-order sideband, and the green lines for the second-order sideband. \textbf{(b)} Time-resolved evolution of photoemission intensity from the first- (solid red line) and second-order sidebands (solid green line). The green dashed line represents the square of the time-resolved evolution of the first-order sideband. \textbf{(c)} Experimentally measured (dots) and simulated (solid lines) pump polarization dependence of the momentum-integrated Volkov sideband intensity, for the first- (red) and second-order (green) sidebands. The dielectric constant was fitted using the polarization dependence of the photoemission intensity from the first sideband (red curve). \textbf{(d)} Experimentally measured (in dotted lines) and theoretical (in solid lines) momentum distribution curves (MDCs) extracted from the three regions of interest shown in (a). The theoretical MDCs are defined as VB$\times|\alpha|^2$ for the first-order (red) and as VB$\times|\alpha|^4$ for the second-order (green) sideband; the MDC of the VB is in black solid line.}
    \label{fig:WSe2}
\end{figure}

To this end, we focus on the investigation of light-dressed 2H-WSe$_2$, using a significantly higher pump fluence ($\sim 5.7$~mJ/cm$^2$). This is possible because, due to the superior surface quality on the spatial scale of our pump beam footprint, we can employ a higher fluence before reaching the limit set by pump-induced space-charge from multiphoton photoemission at defects, compared to GeS and SnS. For these measurements, the incidence angle is still $\theta=65^\circ$, and the incidence plane is along the crystal mirror plane ($\Gamma$-M). The energy–momentum distribution along K–$\Gamma$–K$^{\prime}$ (perpendicular to the incidence plane), at the pump–probe temporal overlap, for a p-polarized probe and integrated over all pump polarizations, is shown in Fig.~\ref{fig:WSe2}(a). In addition to the expected photoemission intensity from the VB, the distribution clearly reveals the emergence of first- and second-order sidebands. 

In Fig.~\ref{fig:WSe2}(b), we examine the temporal dynamics of the photoemission intensity from the first- and second-order sidebands by scanning the pump–probe delay. The second-order sideband (green line) exhibits a noticeably shorter temporal dynamics than its first-order counterpart (red line). This behavior is a universal feature of ultrafast nonlinear light–matter interactions: by definition, higher-order processes exhibit a nonlinear dependence on the field strength and are therefore strongly suppressed at the leading and trailing edges of the pulse. This can be further highlighted by squaring the temporal profile of the first-order sideband (dashed green line in Fig.~\ref{fig:WSe2}(b)), which closely matches the temporal profile of the second-order sideband (green line), a typical behavior for second-order nonlinear processes. In hypothetical scenarios where the pump fluence could be increased even further, higher-order replicas would display an even stronger effective nonlinearity, resulting in an even more pronounced temporal gating relative to lower-order processes.

Having established nonlinear temporal gating of the second-order replica, we are now interested in investigating similar behaviors in its polarization dependence. In Fig.~\ref{fig:WSe2}(c), we show the evolution of the momentum-integrated intensity of the first- (red dots) and second-order (green dots) sidebands as a function of the pump polarization. Both sidebands exhibit the same overall dependence on the pump polarization, with intensity maximized for p-polarization and minimized for s-polarization. However, the second-order sideband displays a distribution that is more sharply peaked around p-polarization. To rationalize these observations, we computed the polarization dependence of the Volkov sideband intensity using the Fresnel–Volkov model. Specifically, we employed the fitted dielectric constant of 2H-WSe$_2$ and assumed a second-order nonlinearity for the second-order sideband. It is well known that the $n^\mathrm{th}$ order Volkov sideband intensity $|\alpha_n|$ is proportional to $|J_n(\alpha)|^2$, where $J_n$ is the $n^\mathrm{th}$ Bessel function of the first kind~\cite{bao_floquet-volkov_2025,park_interference_2014,keunecke_electromagnetic_2020,mahmood_selective_2016,gadge_comparative_2026}. By properties of the Bessel functions, $J_\delta(z) \sim \frac{1}{\Gamma(\delta+1)}\left(\frac{z}{2}\right)^\delta$ for $0 < z \ll \sqrt{\delta+1}$, which yields $J_1(\alpha) \sim \alpha/2$ and $J_2(\alpha) \sim \alpha^2/8$. Therefore, using Eq.~\ref{eq:alpha_revised}, we find $|a_1|^2 \sim |J_1(\alpha)|^2 \propto \cos^2\phi$ and $|a_2|^2 \sim |J_2(\alpha)|^2~\propto~\cos^4\phi$. We note that this reproduces Eq.~\ref{eq:a1} with $\beta = 0$ for the intensity of the first sideband. The observed $\cos^2\phi$ dependence of the first sideband and the $\cos^4\phi$ dependence of the second sideband are in excellent agreement with the polarization-dependent experimental data shown in Fig.~\ref{fig:WSe2}(c).

After investigating the effect of nonlinearity on the temporal and polarization dependence of the second-order sideband, we now turn our attention to its role in the momentum-space distribution of photoelectrons. In Fig.~\ref{fig:WSe2}(d), we show the experimentally measured normalized momentum distribution curves (MDCs) along K–$\Gamma$–K$^{\prime}$ for the VB (black) as well as for the first (red dotted line) and second (green dotted line) sidebands; we also show the theoretical MDCs of the first (red solid line) and second (green solid line) sidebands. It can be seen that, moving from the VB to the first sideband, and from the first to the second sideband, the intensity of the normalized MDC progressively increases around $\Gamma$. This anisotropic enhancement of the signal around $\Gamma$ under p-polarized driving was previously observed in GeS (see Fig.~\ref{fig:GeS_BM}). Here, we additionally show that this enhancement becomes even more pronounced for higher-order sidebands. As shown by the theoretical MDCs of Fig.~\ref{fig:WSe2}(d), this behavior is well captured by Eq.~\ref{eq:alpha_revised} when using the fitted dielectric constant of 2H-WSe$_2$ and accounting for the second-order nonlinearity of the process. In conclusion, this section demonstrates that increasing the pump fluence enables the generation of high-order Volkov replicas, which exhibit clear signatures of nonlinear light–matter interactions in their temporal dynamics, polarization dependence, and photoelectrons angular distributions.

\subsection{Multiple temporal Volkov sidebands}

When light impinges on the surface of a transparent material, part of the beam is reflected at the interface, while the remainder is transmitted into the medium. Until now, our analysis has focused on the effects of the incident and reflected beams on electromagnetic dressing of semiconductors. However, as schematically illustrated in Fig.~\ref{fig:figure1}(a), after partial transmission through the sample, the IR pump beam can be reflected at the interface between the backside of the sample and the substrate and subsequently propagates back toward the top sample–vacuum interface. Upon reaching the surface, this returning field can induce additional excitation processes. These higher-order excitations, arising from the returning pump pulses, are delayed in time by an amount determined by the thickness of the sample and its refractive index. A prominent example of this type of effect in trARPES is the observation of time-domain photoexcitation replicas induced by a ramified pump pulse, reported by Ulstrup \textit{et al.}~\cite{ulstrup_ramifications_2015}. In this work, the authors observed a characteristic double-peak structure in the time-resolved increase of the electronic temperature of graphene on 6H-SiC.

\begin{figure}
    \centering
    \includegraphics[width=\linewidth]{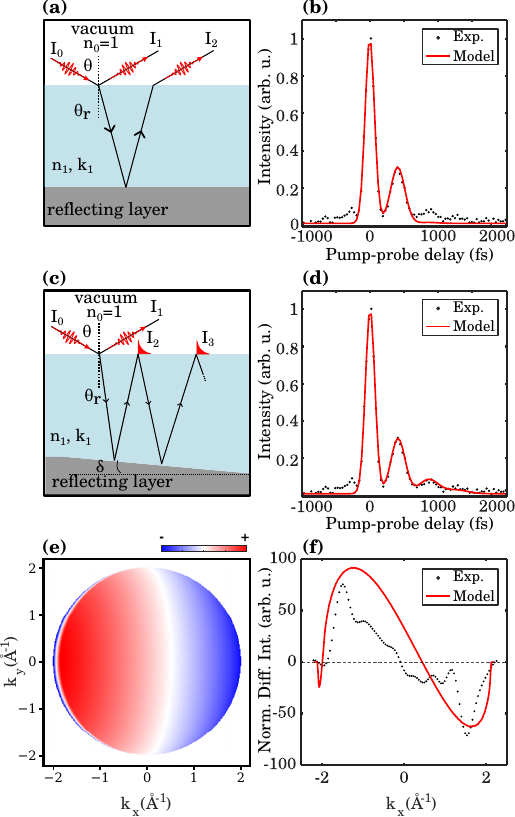}
    \caption{\textbf{Temporally delayed Volkov replicas emerging from multiple reflections inside the material.} \textbf{(a)} Schematic of the multiple reflections of the pump beam inside the GeS sample. At each reflection at the top surface of the material, the beam is partially reflected and partially transmitted. \textbf{(b)} Volkov sideband intensity as a function of pump-probe delay (black dotted line) and associated fit using the simple model based on Fresnel equations and Snell's laws (red solid line). \textbf{(c)} Schematic of the multiple reflections of the pump beam inside GeS, assuming that the sample has a tilt angle $\delta\neq0$ between its top and bottom interfaces, leading to a regime of total internal reflections (TIR, see the main text for more details). Evanescent fields emerge at each reflection at the top surface of the sample, which dress photoelectrons and give rise to the Volkov replicas. \textbf{(d)} Volkov sideband intensity as a function of pump-probe delay (black dotted line) and associated fit using the model taking into account TIR (red solid line). \textbf{(e)} Difference between the momentum-resolved second- and first-order Volkov temporal replica ($I_2$ and $I_1$ in the subpanel~(c)), calculated using the Fresnel-Volkov model. \textbf{(f)} Normalized experimental (black dotted line) and computed (red solid line) differential Volkov sideband intensities integrated along $k_y$. The difference is taken between the second and first peaks ($I_2$ and $I_1$).}
    \label{fig:figure7}
\end{figure}

When scanning the pump–probe delay between the below-band-gap (1.2~eV) pump and XUV probe pulses in GeS, we also observe temporal replicas, in our case in the time-resolved photoemission intensity from the Volkov sideband (Figs.~\ref{fig:figure7}(b) and~(d), black dots). In this section, we aim to provide a unified interpretation of the multiple temporal excitations observed in graphene by Ulstrup \textit{et al.}~\cite{ulstrup_ramifications_2015} and here in GeS, by developing a simple internal reflection model based on geometrical optics combined with Fresnel equations. 

More specifically, the framework developed in this section aims at capturing the evolution of the partially reflected and refracted components of the IR field inside the material and at elucidating how they modulate photoexcitation through successive time-domain interactions. 
Following the schematic representation in Fig.~\ref{fig:figure7}(a), we consider the same IR field as in Fig.~\ref{fig:figure1}(a), incident from vacuum ($n_0=1$) onto a sample with a complex refractive index $\tilde{n}_1$. The beam arrives at an angle of incidence $\theta$, where it undergoes partial reflection and refraction at the vacuum-sample interface. The refracted component propagates through a slab of thickness $d$, reflects at the bottom sample-substrate interface, and returns toward the surface, where it again experiences partial transmission and reflection.

This sequence repeats for each round-trip of the beam inside the material. Each transmitted field component gives rise to a temporally delayed photoexcitation event, \textit{e.g.}, increase of the electronic temperature (graphene~\cite{ulstrup_ramifications_2015}) or Volkov sideband creation (GeS). For p-polarized excitation, the Fresnel coefficients for the power reflection ($R_{01}^\mathrm{p}$) and power transmission ($T_{01}^\mathrm{p}$) read:
\begin{equation}
\begin{aligned}
R_{01}^\mathrm{p} &= 
\left( 
\frac{n_1\cos\theta-\cos\theta_r}
     {n_1\cos\theta+\cos\theta_r} 
\right)^{\!2},
\\[4pt]
T_{01}^\mathrm{p} &= 1 - R_{01}^\mathrm{p}.
\end{aligned}
\end{equation}
Here $\theta$ is the incidence angle with respect to the normal of the sample surface and $\theta_r=\arcsin\left(\sin\theta/n_1\right)$ is the refraction angle (Snell's law).
For simplicity, we omit the superscript $p$ in the following expressions. Energy conservation and reciprocity implies $R_{10} = R_{01}$ and $T_{10} = T_{01}$. We assume partial reflection ($R_{12}$) for the bottom interface, accounting for potential surface roughness and diffuse scattering. We treat $R_{12}$ as a fitting parameter.

By applying Snell’s law, the optical path length $l$ and round-trip time $T$ are:

\begin{align}
    l &= \frac{2d}{\cos\!\left[\arcsin\!\left(\tfrac{1}{n_1}\sin\theta\right)\right]},\label{eq:distance} \\
    T &= \frac{l}{v_g},
    \label{eq:Times}
\end{align}
where $v_g = c / n_1$ is the group velocity of the IR photons inside the slab. The temporal spacing between successive peaks, which is directly accessible in trARPES measurements by varying the pump-probe delay, thus provides a means to extract the sample thickness (assuming that $v_g$ is known).

With this model, we can also express the relative intensity of temporal replicas (see for reference Figs.~\ref{fig:figure7}(b) and~(d)). We assume that the lateral displacement of the pump beam after these round-trips in the crystal remains small compared to the probe beam spot size. We now assume a non-zero absorption, which is accounted for by the imaginary part of the refractive index, $k$. Taking $R_{12}$ as a free parameter that describes incoherent losses at the bottom interface and $\lambda_0$ the wavelength of the incoming IR field, yields the intensity of the $j^{\text{th}}$ peak:

\begin{equation}
\begin{aligned}
I_1 &= I_0\left(1+R_{01}-2\sqrt{R_{01}}\cos2\theta\right), & (j = 1), \\[4pt]
I_j &= T_{01}^2R_{12}^{j-1}R_{10}^{j-2}
       \exp\left(-\tfrac{2(j-1)\pi kl}{\lambda_0}\right)I_0, & (j > 1).
\end{aligned}
\label{eq:noTIR}
\end{equation}
The exponential factor accounts for absorption during each round-trip, following Beer-Lambert attenuation.

To test this simple model based on Fresnel equations and Snell's law, we first benchmark it against the experimental data of Ulstrup \textit{et al.}~\cite{ulstrup_ramifications_2015}, who reported a double time-domain photoexcitation of graphene on 6H-SiC. The agreement between our model and Ref.~\cite{ulstrup_ramifications_2015} is excellent. Indeed, for 6H-SiC at $\lambda_0 = 800\,\mathrm{nm}$ ($\tilde{n}_2 = 2.5 + 0i$~\cite{singh_nonlinear_1971}) and an incidence angle of $\theta = 45^\circ$, assuming negligible absorption ($k = 0$) and perfect reflection at the bottom interface ($R_{12} = 1$), we predict an intensity ratio of $I_2/I_1 = 0.71$. This is in close agreement with the experimentally reported value of $I_2/I_1\sim0.8$. Moreover, using the time delay between the two excitation processes, we retrieved a substrate thickness of $d_\mathrm{theo} = 391\,\mu\mathrm{m}$ compared to the experimental value of $d_\mathrm{exp} = 390\,\mu\mathrm{m}$.

We then apply the same model to fit our experimental pump-probe data on GeS, which shows the emergence of multiple temporal Volkov sidebands. In our experiment, the GeS single crystal is pumped along the zigzag direction under the experimental geometry shown in Fig.~\ref{fig:figure1}(a), with both p-polarized IR pump and XUV probe beams. The time-resolved (momentum-integrated) photoemission signal from Volkov sideband is shown in Fig.~\ref{fig:figure7}(b).
As discussed in section~\ref{sec:alpha}, accurately accounting for all electric field components at the surface is crucial to correctly describe the Volkov intensity and its momentum distribution. The Volkov intensity is proportional to $|\alpha|^2$, which depends on the $E_x$, $E_y$, and $E_z$ components of the electric field, each of which is determined by the angle of incidence $\theta$, the polarization angle $\phi$, and the dielectric constant $\epsilon$.
Therefore, to better model the relative peak intensities, we calculate the electric field components at the surface for each round-trip using Eq.~\ref{eq:Etot}. 
We then use these calculated quantities to retrieve the momentum-integrated $|\alpha|^2$ for each peak $j$, namely:
\begin{equation}
    \mathcal{A}_j = \int_k\, \alpha^2(\textbf{k})\ \mathrm{d}\textbf{k}.
\end{equation}
Then to retrieve the momentum-integrated Volkov sideband intensity, we simply multiply the field intensity at the surface from Eq.~\ref{eq:noTIR} with the retrieved momentum-integrated $\alpha^2$, \textit{i.e.}, $\mathcal{A}_j$:
\begin{equation}
    I_j^\mathrm{Volkov}=I_j\,\mathcal{A}_j.
\end{equation}
We can then use this equation to fit the experimental Volkov intensity shown in Fig.~\ref{fig:figure7}(b). The experimental data are shown as black dots and the model prediction in red line. For GeS, we use $\tilde{n}_2 = 3.8 + ik$ at $\lambda_0=1030$~nm~\cite{al-basheer_determining_2024}, treating $k$ and $R_{12}$ as free fitting parameters. The thickness of the sample is not a fitting parameter as it can be retrieved using Eq.~\ref{eq:Times} and the experimentally accessible round-trip time $T$. A least-squares minimization is performed using the three linearly independent observables $I_2$, $I_3$, and $I_4$, after normalizing to the first peak intensity $I_1$ (see Fig.~\ref{fig:figure7}).

As shown in Fig.~\ref{fig:figure7}(b), we experimentally measured sidebands up to the fourth temporal order (the third order is clearly visible, and the fourth order is visible but faint), whereas the model predicts that only the first two peaks have appreciable intensity. This simplified model drastically underestimates the contribution from temporal replicas beyond the second order. This discrepancy can be understood by noting that, within the model, propagation from the high-index sample into the low-index vacuum results in the beam being almost completely transmitted out of the sample. Consequently, temporal replicas beyond the second order are strongly suppressed.

Since this simplified model fails to reproduce our experimental observations, it calls for refinement through the inclusion of additional physical ingredients. Two uncaptured experimental observations can help identify the missing ingredients in the original model. First, the temporal delays between the first and second (408~fs), and the second and third replicas (443~fs) are not the same: they differ by 35~fs. In addition, the normalized intensity of the third replica ($I_3$) is equal to the square of the normalized intensity of the second replica ($I_2$), \textit{i.e.}, $I_3 \sim (I_2)^2$.

As explained above, these observations are not consistent with the simple model based on Fresnel and Snell's laws. First, the different temporal delays between the first and second, and the second and third replicas cannot be captured by a model that assumes perfectly parallel top and bottom interfaces. Second, the model is also incoherent with the intensity ratio of $I_3 \sim (I_2)^2$. Indeed, Eq.~\ref{eq:noTIR} predicts $I_3 \neq I_2^2$ due to the $R_{10}^{j-2}$ term from partial reflection and refraction on the top surface. 

Having this in mind, we found that the inclusion of a small tilt angle $\delta$ between the top and bottom interfaces was the key ingredient to capture the experimental observations. This tilt angle allows total internal reflections (TIRs) to occur, trapping IR photons within the slab, as sketched in Fig.~\ref{fig:figure7}(c). The tilt angle also explains the different optical paths between the multiple round-trips inside the material, leading to the different temporal delays between the first and second, and the second and third replicas. Moreover, being in the regime of TIR eliminates the $R_{01}^{j-2}$ term by making the top surface perfectly reflective ($R_{10}=1$), restoring the observed $I_3 \sim (I_2)^2$ relation through pure absorption losses. 

Entering the TIR regime due to the small tilt $\delta$ implies that:
\begin{equation}
    \sin(\theta_r + \delta) > \frac{1}{n_1}.
\end{equation}
The introduction of this small tilt angle $\delta$ thus simultaneously satisfies the progressively longer photon paths for higher replicas and entering the TIR regime, since it brings the new incidence angle on the top surface beyond the limit angle: $\theta_r+\delta>\theta_\mathrm{max}$. 
Assuming perfect reflectivity at the top surface ($R_{10}=1$), the only remaining losses are due to absorption encoded in the imaginary part of the refractive index $k$ and the imperfect reflectivity $R_{12}$ at the tilted bottom interface.

Being in the regime of TIR following the penetration of the beam inside the material implies perfect reflectivity at the top surface ($R_{10} = 1$) after the first reflection at the bottom surface of the sample. Therefore, the pulse reaching the interface is not coupled out into the vacuum. This naturally raises the question: how can Volkov sidebands emerge from a field that is trapped inside the material by TIR? The answer lies in the generation of an evanescent field at the interface upon TIR, which is a very well-known effect. This non-propagating evanescent electromagnetic field extends a short distance into the less-dense medium (vacuum), decays exponentially with distance, and can interact with nearby matter (such as the outgoing photoelectrons). The use of evanescent fields in surface-sensitive spectroscopic techniques is well established in the literature. One notable example is the use of attenuated total reflection (ATR) in spectroscopy, which uses the interaction between evanescent fields with matter sitting in the less-dense media, in the vicinity of the interface, to probe their electronic and optical properties~\cite{pepper_optical_1970, neivandt_polarized_1998, holmgren_polarized_2008}. Exploiting the limited path length of evanescent fields, for example, in infrared spectroscopy, avoids the problem of strong attenuation of the photonic signal in highly absorbing media such as aqueous solutions. Furthermore, in 1969, Pepper~\cite{pepper_enhanced_1969} demonstrated that ATR can enhance photoemission, providing clear evidence that evanescent fields can be used to drive photoemission processes.

Evanescent waves exhibit different surface electric field components compared to propagating waves. Hence, to correctly estimate the Volkov sidebands intensities in this new regime, we need to assess the evanescent field components and recalculate $\mathcal{A}_j$, correctly accounting for the reshaping of the electric field components due to the tilt angle $\delta$ and TIR.
Using the reflection geometry, the components of the evanescent electric field at the material surface can be obtained through the following expressions~\cite{neivandt_polarized_1998}:
\begin{align}
E_x &= 
\frac{2\cos\theta_i \, \sqrt{\sin^2\theta_i - n_{01}^2}}
{\sqrt{1 - n_{01}^2} \, \sqrt{(1 + n_{01}^2)\sin^2\theta_i - n_{01}^2}}\, \cos{\phi},\label{eq:Exyz_x} \\[8pt]
E_y &= 
\frac{2\cos\theta_i}{\sqrt{1 - n_{01}^2}}\, \sin{\phi},\label{eq:Exyz_y} \\[8pt]
E_z &= 
\frac{2\sin\theta_i \cos\theta_i}
{\sqrt{1 - n_{01}^2} \, \sqrt{(1 + n_{01}^2)\sin^2\theta_i - n_{01}^2}}\, \cos{\phi}.
\label{eq:Exyz_z}
\end{align}

Here \(n_{01} = 1/n_1\), and \(\theta_i\) is the incident angle just under the material top surface, which depends on both the refraction angle \(\theta_r\) and the incremental tilt \(\delta\) acquired at each reflection from the bottom interface. The parameter \(\phi\) denotes the polarization angle of the incoming IR field as in Fig.~\ref{fig:field_component}(a).
The total field \textbf{E} associated with these field components ($E_x$, $E_y$, $E_z$), can be used to compute the momentum-integrated Volkov sideband intensity $\mathcal{A}_j$.

Including this quantity into Eq.~\ref{eq:noTIR} and imposing the TIR condition $R_{10}=1$, we can rewrite the intensity of the $j^{\text{th}}$ Volkov sideband:
\begin{equation}
    \begin{aligned}
I_1 &= I_0\left(1+R_{01}-2\sqrt{R_{01}}\cos2\theta\right)\mathcal{A}_1,\,&(j=1), \\[4pt]
I_j &= T_{01}\, R_{12}^{\,j-1}\,
       \exp\!\left(-\tfrac{2\pi k\, l_j}{\lambda_0}\right)\mathcal{A}_jI_0,&(j>1),
\end{aligned}
\label{eq:TIR}
\end{equation}
where $l_j$ is the photon path length inside the slab during the $j^{\text{th}}$ roundtrip.

When the interface is tilted, $l_j$ can be expressed via the recursive relation:
\begin{equation}
    l_j = l_{j-1} + \frac{d_j}{\cos\!\left(\theta_r + (j-1)\delta\right)}
                 + \frac{d_j}{\cos\!\left(\theta_r + j\delta\right)},
\end{equation}
where $d_j$ is the local slab thickness at which the field reaches the substrate during the $j^{\text{th}}$ pass. Although both $\delta$ and $d_j$ influence the optical path, $l_j$ is independently constrained by the measured round-trip times $T_j$ through Eq.~\ref{eq:Times}, which remains valid.

In Fig.~\ref{fig:figure7}(d), we compare the experimental Volkov sidebands' intensities (black dots) with a least-squares fit of the revised model (red curve), using $\{k,\, R_{12},\, \delta\}$ as free parameters.
Given the relative complexity of the model, the fitting follows a step-wise procedure to constrain one parameter at a time. First, the sample thickness $d_1$ is determined from the first round-trip time inside the material. The next two round-trip times ($T_2$ and $T_3$) are then used to extract the sample tilt $\delta$. Finally, with $d_1$ and $\delta$ fixed, $k$ and $R_{12}$ are retrieved by fitting the peak intensities using Eq.~\ref{eq:TIR}.
The model reproduces all observed peak intensities and relative delays with excellent quantitative agreement. The best fit values obtained from the least-squares minimization are:
\begin{align*}
\delta &= 1.50^\circ \pm 0.06^\circ, \\
k &= 2.4 \times 10^{-9} \pm 1.6 \times 10^{-3}, \\
R_{12} &= 0.27 \pm 0.17.
\end{align*}

These values are in agreement with our assumption of very weak absorption, consistent with the transparency region of GeS~\cite{al-basheer_determining_2024}, reflect the low reflectivity of the bottom interface (maybe due to a high surface roughness of the silver paste used to glue the sample on the sample holder) and yield a realistically small tilt angle between the top and the bottom surface of the sample. The displacement of the pump beam at the sample surface after its second round trip inside the material can be estimated using simple trigonometry and Eq.~\ref{eq:distance}, yielding a value of approximately $15~\mu\text{m}$. Given the sub-$50~\mu\text{m}$ spot size of the probe beam and the significantly larger spot size of the pump beam ($139~\mu\text{m}\times70~\mu\text{m}$~\cite{Fragkos2025}), the spatial overlap between the probe and the evanescent field is preserved~\cite{Fragkos2025}. This spatial overlap is in the order of 90.5\% at the third round-trip of the beam inside the material. Our assumption of sufficient spatial overlap between the XUV beam and the multiple evanescent fields, therefore, remains valid.

These experimental observations, together with the good agreement with our model, demonstrate that under very specific conditions, evanescent fields associated with total internal reflection can play a crucial role in laser-assisted photoemission. This effect is, however, highly specific to our experimental configuration, which involves (i) a pump frequency within the transparency window of the semiconductor, (ii) a sample thickness that enables temporal separation of the different Volkov orders, and (iii) a large angle of incidence, which is typical for momentum microscope detection schemes.

\subsection{Momentum-space reshaping of Volkov sidebands}

A final consideration can be made by combining the insights from the static Fresnel–Volkov framework with those of the time-domain multiple-reflection model introduced above, through an analysis of the momentum distribution of the temporally delayed Volkov sideband intensities. Indeed, since the electric field components differ for successive Volkov sidebands in the time domain, their momentum-space distributions should carry signatures of this field reshaping that drives the electromagnetic dressing.

The connection to the momentum-resolved Volkov sideband intensity directly follows from Eq.~\ref{eq:alpha_revised} of the Fresnel–Volkov model, with \(\mathbf{E}\) taken from Eq.~\ref{eq:Etot} for the first reflection and \(\mathbf{E}\) obtained from Eqs.~\ref{eq:Exyz_x},~\ref{eq:Exyz_y} and~\ref{eq:Exyz_z} for subsequent round-trips. To ensure consistency, the field components are normalized such that  
$|E_x|^2 + |E_y|^2 + |E_z|^2~=~1$.

Fig.~\ref{fig:figure7}(e) displays the calculated differential momentum-resolved intensity $\Delta I (k_x, k_y)$ between the first and second Volkov sidebands:
\begin{equation}
\Delta I (k_x, k_y) = I_2(k_x, k_y)-I_1(k_x, k_y),
\end{equation}
which is directly proportional to $|\alpha|^2$ for each temporal replica.
In Fig.~\ref{fig:figure7}(e), red (positive differential signal) corresponds to higher photoemission intensity in the second replica, while blue (negative differential signal) corresponds to higher intensity in the first replica within these specific momentum-space regions. The pronounced change in the angular distribution from the first to the second sideband highlights how the transition from propagating fields (first pass) to evanescent fields (subsequent TIR-induced round-trips) leaves a clear fingerprint on the momentum structure of the emitted Volkov sidebands.

This point is further illustrated in Fig.~\ref{fig:figure7}(f), which shows the differential photoemission intensity integrated along $k_y$: the simulated distribution (red) is compared with the experimental data (black, normalized and Fourier-filtered). Both simulation and experiment reveal a reshaping of the momentum distribution, characterized by an enhanced photoemission intensity in the backward hemisphere for the second-order temporal Volkov sideband compared to the first, reflecting the different field components responsible for generating the sidebands. This differential photoemission intensity also accounts for the different orientation of the electric field between the first and delayed sidebands, which occurs because of the tilt of the sample involved in the $\theta_i$ parameter of Eqs.~\ref{eq:Exyz_x},~\ref{eq:Exyz_y} and~\ref{eq:Exyz_z}. The qualitative agreement between experiment and simulation confirms that the combined Fresnel–Volkov and multiple-reflection models accurately capture even subtle features, such as the momentum dependence of the Volkov signal.

\section{Conclusions}

Now that trARPES has become a well-established and powerful technique for probing Floquet physics in solids, obtaining a quantitative understanding and description of inherently competing Volkov transitions is of central importance. In this context, Volkov dressing in metallic systems has been recently investigated in details~\cite{keunecke_electromagnetic_2020,wenthaus_insights_2024}. In metals, it was demonstrated that short electromagnetic penetration depths and strong electronic screening hinder the formation of Floquet-Bloch states, making the Volkov dressing mechanism dominant. By examining how strong screening, shallow penetration depths, and the redistribution of driving fields at metal interfaces shape Volkov sidebands, this body of work has highlighted open questions concerning materials with weaker electronic screening.

In this work, we quantitatively investigate Volkov dressing in the layered semiconductors GeS, SnS, and 2H-WSe$_2$, where moderate screening and the associated deeper penetration of the driving field enable the exploration of new physical phenomena arising from electromagnetic dressing. By employing a quantitative model to describe the polarization-dependent Volkov sideband intensity measured under below-band-gap excitation, we extract lower bound values for the real part of the dielectric function of GeS and SnS. These values lie between the reported monolayer and bulk limits, reflecting the surface sensitivity of trARPES measurements.

To investigate more subtle features related to dielectric screening in the electromagnetic dressing of semiconductors, we employed the revised Fresnel-Volkov model to quantitatively describe the polarization- and momentum-resolved Volkov intensities, which are also sensitive to the materials' dielectric properties. We showed that this model accurately captures the experimentally measured polarization-dependent forward/backward and up/down asymmetries in momentum-resolved Volkov signals.

At higher pump fluences, which are typically required for efficient Floquet states formation, higher-order Volkov sidebands emerge. The nonlinear processes underlying the generation of these high-order sidebands leave characteristic signatures in their temporal dynamics, polarization dependence, and momentum distributions. All of these experimentally observed features were consistently reproduced by incorporating the relevant nonlinearities into the model previously described.

Finally, we exploited the fact that below-band-gap pumping occurs within the transparency window of semiconductors to investigate intriguing aspects related to the time-domain dynamics of Volkov dressing. Our pump-probe measurements reveal multiple temporally delayed Volkov replicas. We showed that these high-order temporal replicas arise from electromagnetic dressing with the evanescent field associated with multiple total internal reflections of the driving field within the crystal. We extended the previously described model to include electromagnetic dressing by evanescent fields. This extended model captures both the temporal delays and the relative intensities of these delayed Volkov sidebands, enabling the extraction of the imaginary part of the dielectric function. In addition, it accurately reproduces the reshaping of the Volkov sideband momentum distribution between the direct and temporally delayed replicas.

Our study reveals several previously unexplored aspects of Volkov dressing in semiconductors and highlights the central role of dielectric screening in shaping the response to the driving field. However, in our analysis, the contributions of Floquet-Bloch states and the resulting Floquet-Volkov interferences were neglected, representing the main limitation of the current model. In the spirit of several recent works in the field~\cite{fragkos_floquet-bloch_2025, merboldt_observation_2025, choi_observation_2025, bao_floquet-volkov_2025, gadge_comparative_2026}, the development of a general unified and quantitative theoretical framework that captures both Floquet and Volkov transitions in semiconductors, where the driving field is partially screened and reshaped near the surface, constitutes an obvious and important direction for future work.

\section{Acknowledgements} We acknowledge Michael Schüler for stimulating discussions. We thank Baptiste Fabre for implementing and maintaining the data binning code. We thank Nikita Fedorov, Romain Delos, Pierre Hericourt, Rodrigue Bouillaud, Laurent Merzeau, and Frank Blais for technical assistance. We acknowledge the financial support of the IdEx University of Bordeaux/Grand Research Program ``GPR LIGHT". This work is part of the ULTRAFAST and TORNADO projects of PEPR LUMA and was supported by the French National Research Agency, as a part of the France 2030 program, under grants ANR-23-EXLU-0002 and ANR-23-EXLU-0004. We acknowledge support from ERC Starting Grant ERC-2022-STG No.101076639, Quantum Matter Bordeaux, AAP CNRS Tremplin, and AAP SMR from Université de Bordeaux. S.F. acknowledges funding from the European Union’s Horizon Europe research and innovation programme under the Marie Skłodowska-Curie 2024 Postdoctoral Fellowship No 101198277 (TopQMat). Q.C. acknowledges funding from the TERAQUANTUM project of the Région Nouvelle-Aquitaine. Funded by the European Union. U.D. acknowledges funding from the European Union’s Horizon Europe research and innovation programme under the Marie Skłodowska-Curie grant No HORIZON-MSCA-2023-DN-01 101169225 - SPARKLE. Views and opinions expressed are however those of the author(s) only and do not necessarily reflect those of the European Union. Neither the European Union nor the granting authority can be held responsible for them. 

\begin{center}
\textbf{DATA AVAILABILITY} 
\end{center}

The data that support the findings of this article are openly available on Zenodo~\cite{courtade_2026_18328301}.


\providecommand{\noopsort}[1]{}\providecommand{\singleletter}[1]{#1}%

\end{document}